# The WARPS survey for faint Clusters of Galaxies

L.R. Jones[1], C.A. Scharf[1], E. Perlman[1], H. Ebeling[2], G. Wegner[3], and M. Malkan[4]

[1] Code 660.2, NASA/Goddard Space Flight Center, MD 20771, USA.
[2] Institute of Astronomy, Madingley Rd, Cambridge, UK.
[3] Dept. Physics & Astronomy, Dartmouth College, 6127 Wilder Lab, Hanover, NH 03755, USA.
[4] Dept. of Astronomy, UCLA, Los Angeles, CA 90024, USA.

**Abstract.** The WARPS cluster survey is based on the ROSAT PSPC archive of pointed observations. It includes extended X-ray sources, detected with the Voronoi Tessellation and Percolation algorithm (VTP), and point-like X-ray sources with non-stellar optical counterparts. It is designed to minimize selection effects whilst covering a large area of sky. The aims of the survey are to (a) measure the low luminosity ($<10^{44}$ erg s$^{-1}$), high redshift ($z>0.2$) X-ray luminosity function of clusters and groups, and (b) investigate cluster morphologies and unusual systems (eg merging clusters). In an initial 13 sq deg (66 fields) we have found 22 extended X-ray sources with detected flux $>7 \times 10^{-14}$ erg cm$^{-2}$ s$^{-1}$ (0.5-2 keV) and sizes of 1 to 5 arcmin. Optically, they range from a single bright nearby galaxy which has been resolved, an Abell cluster which is revealed to have two (probably merging) components, and groups and clusters of galaxies at estimated redshifts beyond $z=0.4$.

## 1. Introduction

Understanding the formation and evolution of clusters of galaxies is of fundamental importance to the study of the growth of structure in the Universe. Complete samples of clusters and groups with which to make such studies can be most reliably constructed using X-ray detection techniques, avoiding the projection effects and selection effects inherent in optical surveys. Current surveys indicate no evolution of the X-ray luminosity function up to redshifts of $z=0.2$ (Ebeling et al, these proceedings), and negative evolution at higher redshifts (Henry et al 1992). However, the cluster X-ray luminosity function (XLF) at luminosities lower than $\approx 10^{44}$ erg s$^{-1}$ has not been measured at $z>0.15$ using a complete, X-ray selected sample. The WARPS (Wide Angle ROSAT Pointed Survey) cluster survey is designed to make this measurement, testing whether low luminosity clusters evolve negatively or positively, and constraining models of the growth of structure in the Universe.

## 2. The WARPS Survey strategy

The WARPS cluster survey strategy is to cover a large area of sky at a flux limit designed to sample a new part of redshift-luminosity space, whilst minimizing selection effects which would bias the survey against X-ray emission from any hot gas associated with a galaxy or galaxies. All sources, both extended and point-like, were initially included in the sample. Results from the VTP source detection algorithm as well as the WGACAT catalogue (White et al 1995) were used. APM machine measurements of POSS and UKST/SERC plates were then used to investigate the contents of the 15 arcsec radius X-ray error circles and correct for the small PSPC boresight position error. CCD imaging at 1m class telescopes (Lick 1m; MDM 1.3m; CTIO 0.9m) was obtained of blank error circles or where the brightest candidate counterpart was within 1 mag of the plate limit, and thus had unreliable star/galaxy classification. Point X-ray sources which had a clear stellar counterpart were then removed from the sample. These are expected to be QSOs and stars. Spectroscopy of the remaining sources has being performed at the Lick 3m, MDM 2.4m and KPNO 4m telescopes; these data are in the process of being analysed.

## 3. The VTP algorithm

The VTP algorithm is a general method for the detection of non-Poissonian structure in a distribution of points (Ebeling & Wiedenmann 1993). Briefly, Voronoi cells are constructed around each raw photon. The reciprocal of the cell size corresponds to a measure of local surface brightness. Neighbouring photons in cells significantly smaller than a threshold above the background are collected together into sources. The background is then recomputed and significant photons redetermined in an iterative procedure. *All* significant sources, point-like and extended, are found by VTP, with no binning or smoothing size required. VTP is particularly suited to finding low surface brightness extended sources. We currently run VTP at three thresholds above the background, and use the higher thresholds where blending of close sources is a problem. We use the observed source size and the position dependent psf to estimate the true source size, and recover the total source flux assuming a King profile. Known clusters at redshifts of z=0.13 (Pavo, Griffiths et al 1992), 0.275 (J1836.10RC), 0.56 (J1888.16CL) and 0.664 (F1767.10TC, all from Couch et al 1991) have all been detected as extended X-ray sources.

## 4. Results

In an initial search of the inner 15 arcmin radius of 66 randomly selected ROSAT PSPC fields (13 sq deg) with exposures >10 ksec at $|b|>20^\circ$, 22 serendipitous extended sources with detected flux $>7\times10^{-14}$ erg cm$^{-2}$ s$^{-1}$ (0.5-2 keV) were found. This flux limit gave signal/noise values of 7-20. The size of the X-ray sources vary from 1 to 5 arcmin in their longest dimension. Simulations show that we are complete at this flux limit for a reasonable range of cluster sizes (core radius <500 kpc at z>0.15). The simulations also show that at our typical surface brightness limit, we detect >50% of the total flux of a cluster at the flux limit of the survey with a King profile of $r_c = 250$ kpc, $\beta = 0.66$ and redshift z>0.1.

Initial spectroscopy has shown that WARPS contains, as expected, some low redshift AGN, including narrow and broad line objects. In these objects the host galaxy is resolved optically, and so the source has remained in the sample. The remaining objects, are, however, low redshift normal galaxies (mostly ellipticals) and clusters and groups of galaxies of a wide range of richnesses and redshifts. One cluster is Abell 2465, which is detected as two separate extended X-ray sources at the same redshift of z=0.245. The two components are separated by $1.5h^{-1}$ Mpc and may be in the process of merging. Several of the highest redshift clusters (at estimated redshifts z>0.5) are point X-ray sources identified as clusters via optical imaging. Thus a combination of both extended and point-like X-ray sources is required to ensure a complete cluster sample, at least in the ROSAT fields studied here.

Further optical follow-ups and extensions to more ROSAT fields and fainter fluxes ($4\times10^{-14}$ erg cm$^{-2}$ s$^{-1}$) are in progress.

## 5. Discussion

The ROSAT PSPC images have sufficient signal/noise that a much lower limiting surface brightness can be reached than in the EMSS (Henry et al 1992). Thus a better estimate of the total flux of sources at the flux limit of the EMSS is obtained, as well as the ability to probe to fainter limits. The VTP algorithm fully exploits this signal/noise in an automated source search. Algorithms based on the instrument psf would underestimate the flux of extended sources, or miss the low surface brightness sources entirely.

The number of sources detected here is slightly higher than that predicted by the LogN-LogS relations in Rosati et al. (1995), although a complete analysis is still underway. The main conclusions of this paper are that (a) large numbers of clusters and groups of galaxies are detectable as extended X-ray sources in the ROSAT PSPC archive, (b) VTP provides an excellent way of both finding them and measuring their fluxes accurately and (c) point-like X-ray sources need to be included in order to be complete at the highest redshifts.

*Acknowledgements.* We are grateful to Mike Irwin for searches of the APM database and Nick White for discussions related to WGACAT.

## References

Couch, W. et al 1991. Mon. Not. R. Ast. Soc., 249, 606.
Ebeling, H. et al 1995, these proceedings.
Ebeling, H. & Wiedenmann, G. 1993. Phys Rev 47, 704.
Griffiths et al 1992. Mon. Not. R. Ast. Soc., 255, 545.
Henry, J.P. et al 1992. ApJ, 386, 408.
Rosati, P. et al 1995. ApJ, 445, 11.
White, N.E., Giommi, P., Angellini, L. HEAD Meeting Nov 1994.
(http://legacy.gsfc.nasa.gov/ white/wgacat/head.html).

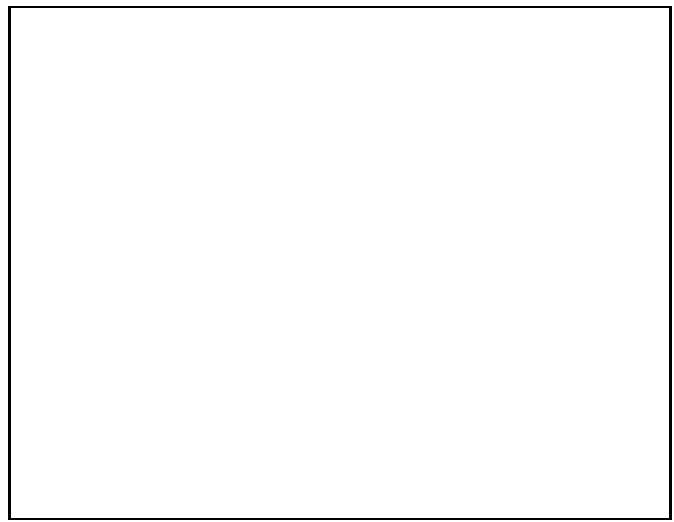

**Fig. 1.** Example of an extended X-ray source (contours) overlaid on an optical CCD R band image. This is a poor group of galaxies of size $\approx 2$ arcmin.